\documentclass[aps,prd,twocolumn,reprint]{revtex4-2}

\usepackage{graphicx}
\usepackage{bm}
\usepackage{hyperref}
\usepackage{amsfonts}
\usepackage{tikz}
\usepackage{amssymb}
\usepackage{graphicx}
\usepackage{color}
\usepackage{epsfig}
\usepackage{remreset}
\usepackage{amsmath,amsthm,ulem}
\usepackage{mathtools}
\usepackage{mathrsfs}
\usepackage{color}
\usepackage{float}
\usepackage{tensor}
\usepackage{lipsum}
\usepackage[caption=false]{subfig}
\usepackage{slashed}
\usepackage{lmodern}
\usepackage[T1]{fontenc}

\begin{document}

\title{The shadow and quasinormal modes of the asymptotically flat hairy black holes with a dilaton potential}

\author{Sheng-Han Xiong}
\email{xsh050113@hotmail.com}
 \affiliation{School of Science, Jiangsu University of Science and Technology, Zhenjiang 212100, Chian}
\author{Xiao-Mei Kuang}%
 \email{xmeikuang@yzu.edu.cn}
\affiliation{ 
Center for Gravitation and Cosmology, College of Physical Science and Technology, Yangzhou University, Yangzhou 225009, China
}%

\author{Yong-Zhuang Li}
\email{liyongzhuang@just.edu.cn}
\affiliation{%
School of Science, Jiangsu University of Science and Technology, Zhenjiang 212100, Chian
}%

\author{ Jerzy Matyjasek}
\email{jurek@kft.umcs.lublin.pl}
\affiliation{%
Institute of Physics, Maria Curie-Skłodowska University
pl. Marii Curie-Skłodowskiej 1, 20-031 Lublin, Poland
}%

\date{\today}

\begin{abstract}
 In this article, the shadow and the quasinormal modes (QNMs) of an exact asymptotically flat hairy electrically charged black hole solution with a dilaton potential are investigated. Using the {constraint} equation among the integration constant $\eta$ of the gravitational field, the mass $M$, the electric charge $Q$ and the coupling constant $\nu$ between the $U(1)$ field and the dilaton field, we find that the shadow radii, the Lyapunov exponent $\lambda$ and the coordinate angular velocity $\Omega_{c}$ only significantly affected by $\nu$ if the $Q$ is close to the extremal value, especially when $\nu$ approaches to one. Furthermore, the QNMs are numerically computed by using the Hatsuda method and verify with the higher-order WKB approximations with the Pad\'e summation. We find that the QNMs are close to that of the low energy limit of the string theory when $\nu$ is large enough. In the eikonal limit, the real and imaginary parts are proved to be given by $\Omega_{c}$ and $\lambda$, respectively.
\end{abstract}

\maketitle

\section{Introduction}\label{sec:intro}

The no-hair theorem states that the asymptotically flat black holes formed through the gravitational collapse should be described by only a few parameters, i.e., the mass, the angular momentum and the charge \cite{Israel:1967za,Hawking:1971vc,Robinson:1975bv,Mazur:1982db,Bekenstein1995prdr,Mayo1996prd,Broderick:2013rlq,Herdeiro:2015waa,Gurlebeck:2015xpa}. However, this theorem could be violated if one considers the non-trivial matter fields in the black hole spacetime, or the quantum effects near the horizon, see \cite{Greene:1992fw,Zloshchastiev:2004ny,Herdeiro2014prl,Herdeiro:2015tia,Hawking:2016msc,Bousso:2017dny,Delgado:2019prc,Peng:2019nke,Hong:2020miv,Herdeiro:2020xmb,Delgado:2020hwr,Gao:2021ubl,Bakopoulos:2021dry} and references therein. With the ability to observe gravitational wave signals and black hole shadows, the verification of the black hole no-hair theorem can now be conducted through these new observational methods  \cite{LIGOScientific:2016aoc,EventHorizonTelescope:2019dse,EventHorizonTelescope:2022wkp}. Numerous investigations have been made in this direction \footnote{For testing the no-hair theorem in other ways, like the electromagnetic spectrum X-ray observations, see \cite{Bambi:2015ldr,Yagi:2016jml,Ni:2016rhz,Johannsen:2016uoh,Bambi:2016sac,Wang:2018bbr} and references therein}, and it is impractical for us to provide an exhaustive list of all relevant works. In other words, some omissions may inevitably occur. For studies constraining hair-related parameters through black hole shadow observables, we refer the readers to the Refs. \cite{Psaltis:2015uza,Cunha:2015yba,Afrin:2021imp,Khodadi:2021gbc,Tang:2022uwi,Wang:2025ihg,Gulia:2025tvx} for hairy Kerr metric, the nice review \cite{Vagnozzi:2022moj} for many other well-motivated deviations from classical General Relativity (GR) black hole solutions and Refs. \cite{Glampedakis:2023eek,Ye:2023qks,Benavides:2024jyw,Yang:2024utv,Zhao:2024exh,QiQi:2024dwc,Liu:2024wal,Shu:2024tut,Raza:2025pgs,Lim:2025cne,Meng:2025ivb,Li:2025ibp,Lim:2025cne} for new investigations in the past two years. Regarding tests of the black hole no-hair theorem via gravitational waves or quasinormal modes (QNMs), see Refs. \cite{Apostolatos:2009vu,Meidam:2014jpa,Yunes:2016jcc,Cardoso:2016ryw,Dhanpal:2018ufk,Islam:2019dmk,Isi:2019aib,CalderonBustillo:2020rmh,Lei:2023wlt,Yunes:2025xwp}.

Indeed, while the Kerr metric currently stands as the most widely accepted model for astrophysical black holes, present observational precision remains insufficient to rule out a considerable number of well-motivated alternative scenarios as noted in Ref. \cite{Vagnozzi:2022moj}. Moreover, the Kerr family solutions derived from classical GR cannot adequately address all outstanding questions concerning black holes, especially the unification of gravity and quantum mechanics. Searching the black hole configurations in quantum gravity has attracted the attention of many physicists over the past few decades \cite{Maldacena:1996ky,Mohaupt:2000gc,Alexeyev:2020cuq,Bojowald:2020dkb,Mayerson:2020tpn,Zhang:2023yps,Bambi2024}. Among them, dilaton black holes have always been a major focus of these investigations, where the the dilaton field is non-minimally coupled to the Maxwell field, i.e., the Einstein-Maxwell-dilaton (EMD) gravity, which originally arises from the the low-energy string theory limit and investigated in Refs. \cite{Gibbons:1987ps,Garfinkle1991prd,Poletti1994prd,Sheykhi2008prd,Fahim:2021nly} (Hereafter, we will simply refer to such black holes as the GMGHS type solutions).

Recently, an exact asymptotically flat charged hairy black holes in a gravity model with a dilaton and its potential has been developed in Refs. \cite{Anabalon2012jhep,Anabalon:2013qua,Astefanesei2019JHEP,Astefanesei:2019qsg,Astefanesei2021JHEP}, which we will refer to as Anabal\'{o}n-Astefanesei-Mann (AAM) black holes. Such model has the similar Lagrangian with the EMD gravity but different physical properties with the GMGHS solutions. Firstly, the AAM black holes have a well defined extremal limit, and interpolate between Reissner-Nordstr\"{o}m and the GMGHS family. Secondly, a sub-class of thermodynamically and dynamically stable hairy black holes exist \cite{Astefanesei:2019qsg}. Thirdly, the dilaton potential arises naturally in a consistent truncation of $\mathcal{N}=2$ supergravity in four dimensions, extended with vector multiplets and deformed by a dyonic Fayet-Iliopoulos term \cite{Anabalon:2017yhv}.

In this article, we will investigate the shadow, the photon region and the scalar radial perturbations of an asymptotically flat, charged spherically symmetric AAM black hole. 

Many sterling works have focused on the optical appearances of the stringy GMGHS families, including the static, spherically symmetric charged dilaton black holes \cite{Maki:1992up,Younesizadeh:2022czv,Heydari-Fard:2021pjc,Promsiri:2023rez}, the phantom (anti-)dilaton black holes wormholes \cite{Azreg-Ainou:2012ity,Ovgun:2018prw}, a spinning Kaluza-Klein dilaton black hole \cite{Amarilla:2013sj,Amarilla:2015pga}, the Einstein-Maxwell-Dilaton-Axion black holes \cite{Wei:2013kza,Flathmann:2015xia,Soroushfar:2016yea,Javed:2019ynm}, the charged wormholes in EMD theory \cite{Amir:2018szm}, the dual-charged stringy black hole \cite{Kala:2020prt}, the static charged and slowly rotating charged dilaton black holes surrounded with thin disks or a plasma \cite{Heydari-Fard:2020ugv,Javier2023prd}, and so on. 
For AAM black holes, the effect of hair on the deflection angle in the weak field limits has been investigated by using Gauss-Bonnet theorem in Ref. \cite{Javed:2019rrg}. On the other hand, it has been suggested that in the eikonal limit, QNMs of spherically symmetric, asymptotically flat black holes are related to the parameters of the circular null geodesics and the gray-body factor \cite{Cardoso:2008bp,Stefanov:2010xz,Konoplya:2024lir}. 

Again, for GMGHS families, the QNMs have been investigated for different cases, see Ref. \cite{Konoplya:2001ji,Fernando:2003wc} for the normal electrically charged dilaton black hole with arbitrary coupling constant, Ref. \cite{Moderski:2001ru} for evolution of a massless charged scalar field around the electrically charged dilaton black hole with coupling constant one, Ref. \cite{Ferrari:2000ep,Shu:2004fj} for QNMs of the Garfinkle-Horowitz-Strominger black hole, Ref. \cite{Chen:2005rm,Fernando:2008hb} for 2+1 dims dilaton black holes with cosmological constant and its extremal case in Ref. \cite{Fernando:2022wlm}. For QNMs of dilaton-(A)dS black holes, see \cite{Lopez-Ortega:2005obq,Fernando:2016ftj,Konoplya:2022zav}. The QNMs and Hawking radiation in the vicinity of an Einstein-dilaton-Gauss-Bonnet black hole has been investigated in Ref. \cite{Konoplya:2019hml}. For dyon-like and higher dimensions dilatonic black hole, see \cite{Malybayev:2021lfq,Becar:2007hu,Lopez-Ortega:2009jpx}. The precise analytic expressions for the QNMs of scalar and Dirac fields in dilaton black holes with coupling constant one have been given in Ref. \cite{Malik:2024sxv}. For the QNMs of the charged fermions in linear dilaton black holes, see \cite{Sakalli:2021dxd}. For the weakly charged dilaton black holes, see \cite{Brito:2018hjh}. For the QNMs of a stationary axisymmetric EMD-axion black hole, see \cite{Pan:2006af}. Finally, the QNMs of EMD black holes with a scalar hair and the Gibbons-Maeda black holes in EMD theories were investigated in \cite{LuisBlazquez:2020rqp,Rincon:2020pne,Pope:2024ncb}. 

Thus, our present work establishes a comparative framework with GMGHS solutions, and provides new confirmation of the geometric-optical correspondence.

The outline of this work is as follows: in Sec. \ref{sec:reviews} we briefly review the asymptotically flat hairy electrically charged black hole and then we compute the shadow and the photon sphere of such kind of black hole in Sec. \ref{sec:shadows}. The QNMs of the linear scalar perturbations are numerically investigated using the Hatsuda method \cite{Bender1969pr,Sulejmanpasic2018CPC,Hatsuda2020prd}, while  verified with the higher-order WKB approximations with the Pad\'e summation \cite{Matyjasek:2017psv,Matyjasek:2019eeu} in Sec. \ref{sec-qnm}. Finally, some summaries and remarks are made on the last section.

\section{The exact asymptotically flat hairy electrically charged black hole solution}\label{sec:reviews}

In this section we will make a brief review about the asymptotically flat AAM black holes adopted here. Following Refs. \cite{Astefanesei2019JHEP,Astefanesei2021JHEP}, the model consists of the Einstein-Hilbert term, one non-minimally coupled Maxwell-dilaton Lagrangian and a non-trivial dilaton potential, with the action given by
\begin{equation}\label{action}
    S=\frac{1}{2\kappa}\int d^{4}x \sqrt{-g}\left[R-\frac{1}{4}e^{\gamma \phi}F^2-\frac{1}{2}(\nabla\phi)^{2}-V(\phi)\right],
\end{equation}
where $\kappa=1/8\pi$ with $c=G_{N}=\hbar=1$. $\gamma$ indicates the coupling constant between the dilaton field $\phi$ and  the electromagnetic field $F_{\mu\nu}=\partial_{\mu}A_{\nu}-\partial_{\nu}A_{\mu}$. The equations of motion (EOMs) are then expressed as 
\begin{eqnarray}
    R_{\mu\nu}-\frac{1}{2}Rg_{\mu\nu}&=&T_{\mu\nu}^{\phi}+T_{\mu\nu}^{EM},\label{eom1}\\
    \partial_{\mu}\left(\sqrt{-g}e^{\gamma\phi}F^{\mu\nu}\right)&=&0,\label{eom2}\\
    \frac{1}{\sqrt{-g}}\partial_{\mu}\left(\sqrt{-g}g^{\mu\nu}\partial_{\nu}\phi\right)&=&\partial_{\phi}V(\phi)+\frac{1}{4}\gamma e^{\gamma\phi}F^{2};\label{eom3}
\end{eqnarray}
where the stress tensors are defined as 
\begin{eqnarray}
    T_{\mu\nu}^{\phi}&=&\partial_{\mu}\phi\partial_{\nu}\phi-g_{\mu\nu}\left[\frac{1}{2}(\partial_{\alpha}\phi\partial^{\alpha}\phi)^{2}+V(\phi)\right],\\
    T_{\mu\nu}^{EM}&=&e^{\gamma \phi}\left(F_{\mu\alpha}F_{\nu}^{~\alpha}-\frac{1}{4}g_{\mu\nu}F_{\alpha\beta}F^{\alpha\beta}\right).
\end{eqnarray}
The determination of the specific form of the dilaton potential $V(\phi)$ allows for the explicit derivation of the spacetime metric. A general self-interacting potential is
\begin{eqnarray}
    V(\phi)&=&\frac{2\alpha}{\nu^2}\left[\frac{\nu-1}{\nu+2}\sinh{\left(\sqrt{\frac{\nu+1}{\nu-1}}\phi\right)}\right.\nonumber\\
    &~&\qquad\left.-\frac{\nu+1}{\nu-2}\sinh{\left(\sqrt{\frac{\nu-1}{\nu+1}}\phi\right)}\right.\nonumber\\
     &~&\qquad \left.+4\left(\frac{\nu^2-1}{\nu^2-4}\right)\sinh{\left(\frac{\phi}{\nu^2-1}\right)}\right],
\end{eqnarray}
where the real constant $\alpha$ parametrizes the strength of the potential with dimension $[\alpha]=L^{-2}$ and 
\begin{eqnarray}
    \gamma=\sqrt{\frac{\nu+1}{\nu-1}}.
\end{eqnarray}
As well known, $\gamma=0$ corresponds to the classical Einstein-Maxwell theory plus a scalar field. $\gamma=1$ describes the tree-level low energy limit of the string theory and $\gamma=\sqrt{3}$ corresponds to the compactification of the Kaluza-Klein theory. In the article we consider $\gamma>1$, i.e., $\nu\geq 1$. 

The general static, spherically symmetric, asymptotically flat metric has also presented as
\begin{eqnarray}\label{metric}
    ds^{2}&=&\Psi(x)\left[-f(x)dt^2+\frac{\eta^{2}dx^2}{f(x)}+d\Omega_{2}^{2}\right],\\
    A_{\mu}dx^{\mu}&=&-\frac{q}{\nu x^{\nu}}dt,\quad \phi(x)=\sqrt{\nu^2-1}\ln{x},
\end{eqnarray}
where $d\Omega_{2}^{2}=d\theta^2+\sin^{2}{\theta}d\varphi^2$ and $\Psi(x)$, $f(x)$ are 
\begin{eqnarray}
    \Psi(r)&=&\frac{\nu^2 x^{\nu-1}}{\eta^2 (x^{\nu}-1)^{2}},\\
    f(x)&=&\frac{1}{\nu^2}\left[\alpha\left(\frac{x^{\nu+2}}{\nu+2}-x^2+\frac{x^{2-\nu}}{2-\nu}+\frac{\nu^2}{\nu^2-4}\right)\right.\nonumber\\ 
    &~& \left.+\eta^2\left(1-\frac{q^2}{2\nu x^{\nu}}\frac{x^{\nu}-1}{\nu-1}\right)(x^{\nu}-1)^{2}x^{2-\nu}\right].
\end{eqnarray}
$q$ and $\eta$ are two integral constants describing the conserved charges of the solutions. $x$ is the dimensionless radial coordinate \cite{Anabalon2012jhep} and we consider $x\in]1,\infty]$. The condition of asymptotically flat is enforced by requiring 
\begin{equation}
    \lim_{x\rightarrow 1}\Psi(x)f(x)=1.
\end{equation}
Meanwhile, if one rewrites the metric in Schwarzschild coordinates $(t, r, \theta, \varphi)$ then $\lim_{x\rightarrow1}\Psi(x)=r^2+\mathcal{O}(r^{-1})$. In fact, for most asymptotically flat spacetime with the two-sphere $R(r)d\Omega_{2}^{2}$, the transformation $x\leftrightarrow r$ of coordinates can be determined by $\Psi(x)=R(r)$. The ADM mass and the electric charge of the black hole are computed to be
\begin{eqnarray}
    M&=&\frac{q^2}{4\eta(\nu-1)}-\frac{\alpha+3\eta^2}{6\eta^3},\label{eq-mass}\\
    Q&=&-\frac{q}{4\eta}. \label{eq-charge}
\end{eqnarray}
Without loss of generality, we set $M=1$. So by (\ref{eq-mass}, \ref{eq-charge}) and $\nu\geq1$ the parameter $\eta$ will be restricted by $\eta>\eta_{c}$, where $\eta_{c}$ is the solution of $\alpha+3\eta^2+6M\eta^{3}=0$. Note by introducing the transformations 
\begin{eqnarray}
    \eta\rightarrow\sqrt{\alpha}\eta,\quad M\rightarrow\sqrt{\alpha}M,\quad Q\rightarrow\sqrt{\alpha}Q
\end{eqnarray}
$\alpha$ will no longer appear explicitly in the relevant physical expressions. In addition, the relations (\ref{eq-mass}, \ref{eq-charge}) leads to 
\begin{eqnarray}
    \nu=1+\frac{24 Q^2 \eta^4}{\alpha+3\eta^2+6M\eta^3}. \label{eq-nu}
\end{eqnarray}
So for $Q\rightarrow0$ one has $\nu\rightarrow1$, the spacetime corresponds to Schwarzschild black hole; For a finite value of $Q$ and $\eta\rightarrow\infty$ the spacetime corresponds to a charged dialton black hole with $\gamma=1$.

\section{The black hole shadow and the photon sphere}\label{sec:shadows}

In this section we will investigate the null geodesics and the black hole shadows under the current spacetime. Due to the spherical symmetry, the Lagrangian can be then written as 
\begin{equation}
    \mathcal{L}=\frac{1}{2}(g_{tt}\dot{t}^2+g_{xx}\dot{x}^{2}+g_{\varphi\varphi}\dot{\varphi}^2),
\end{equation}
where the dot represents the derivative with respect to an affine parameter $\lambda$ and we have set $\theta=\pi/2, \dot{\theta}=0$. The conserved energy $E$ and orbital angular momentum $L$ per unit mass of the particles are 
\begin{eqnarray}
    E=-g_{tt}\dot{t}, \quad L=g_{\varphi\varphi}\dot{\varphi}.
\end{eqnarray}
The orbital equation of photons approaching to the black hole on the orbital planes is now
\begin{eqnarray}
    \left(\frac{dx}{d\varphi}\right)^{2}=\frac{E^2}{\eta^2L^2}-\frac{f(x)}{\eta^2}.
\end{eqnarray}
The conditions required for obtaining unstable circular null geodesic are thus 
\begin{eqnarray}
    f(x_{ph})&=&\frac{E^2}{L^2}=b^{-2}, \label{eq-cng1} \\
    f'(x_{ph})&=&0, \label{eq-cng2}
\end{eqnarray}
where $b\equiv L/E$ is the impact parameter. Correspondingly, the Lyapunov exponent $\lambda$ and coordinate angular velocity $\Omega_c$ at the unstable null geodesic are defined as below \cite{Cardoso:2008bp}:
\begin{eqnarray}
    \lambda&=&\sqrt{\frac{\Psi^{2}(x)f^{2}(x)}{2}\frac{d^2}{dx^2}\left(\frac{1-b^2f(x)}{\eta^2 \Psi^{2}(x)}\right)}\Bigg\vert_{x\rightarrow x_{ph}}, \label{eq-lambda}\\
    \Omega_{c}&=&bf(x)\vert_{x\rightarrow x_{ph}}.\label{eq-omegac}
\end{eqnarray}
The black hole shadow radius can be calculated in terms of the photon sphere as \cite{Perlick2022PR}
\begin{eqnarray}
    x_{sh}=\frac{x_{ph}}{\sqrt{\Psi(x_{ph})f(x_{ph})}}.
\end{eqnarray}

Meanwhile, to ensure $f(x)=0$ has at least one root other than $x=1$, i.e., the event horizon should exist, $Q,\,\eta,\,\nu$ must satisfy certain constraints. Considering that both $f(x)$ and $f'(x)$ equal to zero as $x$ tends to 1, while $f''(x\rightarrow1)=2\eta^2>0$, this indicates that $f(x)$ is a concave function as $x\rightarrow1$. So the event horizon will exist when $f'(x\rightarrow\infty)<0$. This constraint leads to 
\begin{eqnarray}
    8Q^{2}\eta^4(\nu+2)>(\nu-1)\nu(\alpha+\eta^2(2+\nu)).\label{eq-restricted}
\end{eqnarray}
 
\begin{figure*}
\includegraphics[width=\textwidth]{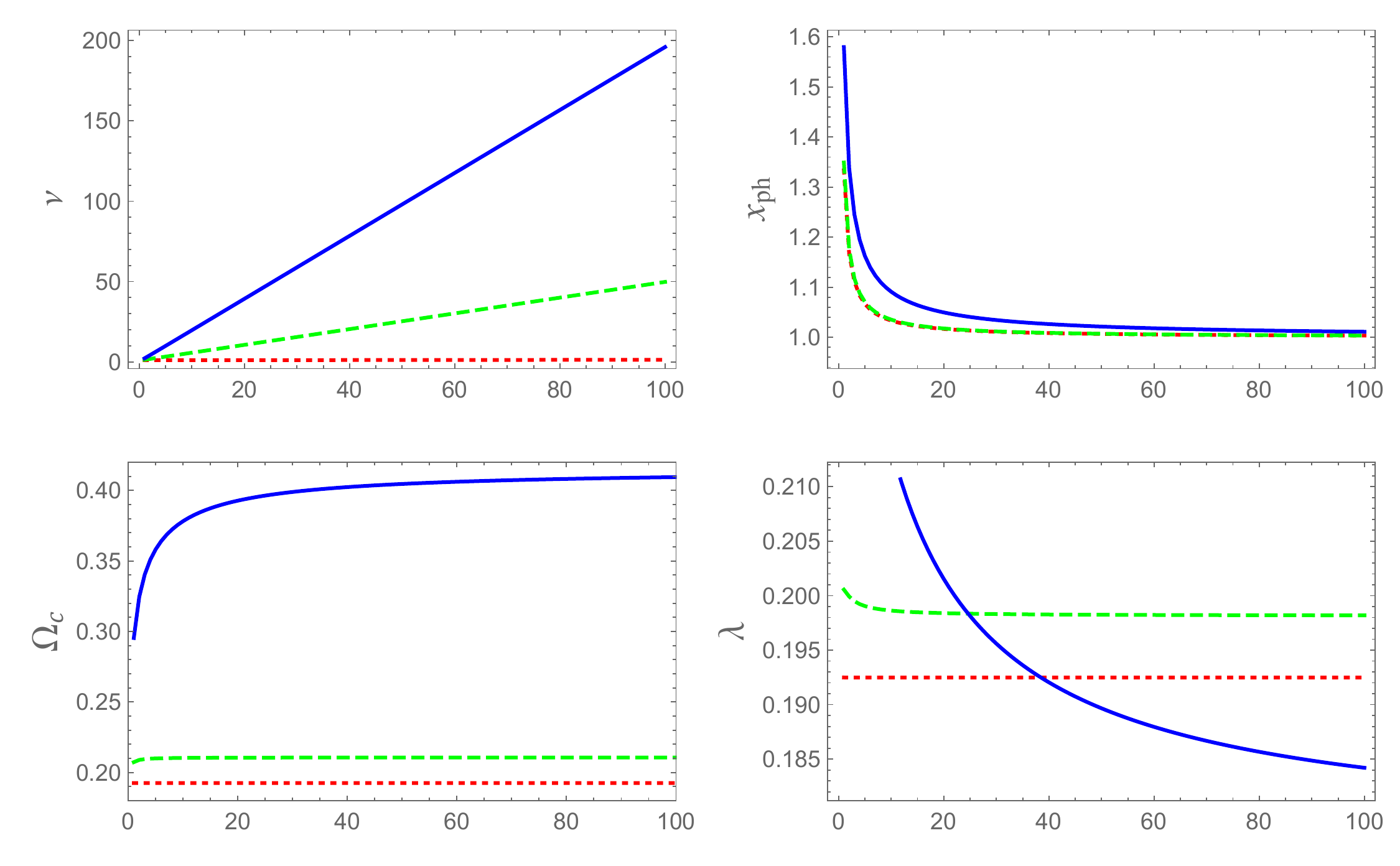}
\caption{\label{fig:eta}$\nu,\,x_{ph},\,\Omega_{c},\,\lambda$ as functions of $\eta$. For all subfigures, the horizontal coordinate is $\eta$. The red dotted line, the green dashed line and the blue solid line are for $Q=0.025,\,0.35,\,0.7$, respectively.}
\end{figure*}

Fig. \ref{fig:eta} shows the parameters $\nu,\,x_{ph},\,\lambda,\,\Omega_{c}$ as functions of $\eta$ with fixed charge $Q$. As restricted by Eq. (\ref{eq-mass}), $\nu$ is almost linear to $\eta$. Therefore, $x_{ph},\,\lambda,\,\Omega_{c}$ as functions of $\nu$ should have similar behaviors as functions of $\eta$. The figures also show that as $\eta$ increasing, $x_{ph},\, \lambda,\,\Omega_c$ all approach to constants. Actually, as $\nu\rightarrow\infty$ the spacetime degrades into the low energy limit of the string theory with $\gamma=1$, which indeed leads to the fixed values for each parameters. Especially, as $Q\rightarrow 0$ the spacetime becomes the Schwarzschild one, where $\lambda^{S}=\Omega^{S}_{c}=27^{-1/2}$ \cite{Perlick2022PR}. Especially, we find that large $Q$ will significantly affect the behavior of the coordinate angular velocity and the Lyapunov exponent.

\section{The linear scalar perturbations}\label{sec-qnm}

We now consider the linear perturbations of the dilaton black hole (\ref{metric}). Usually, there are two ways to perform the perturbations, by adding test fields to the black hole spacetime or by perturbing the metric \cite{Konoplya2011rmd}. For a test massless scalar field $\Phi$ the equation of motion is given by the general covariant Klein-Gordon equation
\begin{eqnarray}\label{eq-scaper}
    \frac{1}{\sqrt{-g}}\partial_{\nu}\left(g^{\mu\nu}\sqrt{-g}\partial_{\mu}\Phi\right)=0,
\end{eqnarray}
where $g_{\mu\nu}$ denotes the background metric. Due to the spherical symmetry of the current spacetime, the wave equation $\Phi(t,x,\theta,\varphi)$ can be expressed as
\begin{equation}\label{eq-wavef}
    \Phi(t,x,\theta,\varphi)=\sum_{l,m}\frac{R(x)}{\sqrt{\Psi(x)}}Y_{l,m}(\theta,\varphi) e^{-i\omega t},
\end{equation}
where $Y_{l,m}(\theta,\varphi)$ are the spherical harmonics. Substituting (\ref{eq-wavef}) into the Klein-Gordon equation (\ref{eq-scaper}), one then obtains a Schr$\text{\"{o}}$dinger-like equation for radial variable
\begin{eqnarray}
    \frac{d^{2}R}{dx_{*}^{2}}+\left[\omega^2-V(x)\right]R=0,
\end{eqnarray}
where $l$ denotes the multipole quantum number arising from the separation of angular variables. $x_{*}$ represents the "tortoise" coordinate, determined by
\begin{eqnarray}
    dx_{*}=\frac{\eta dx}{f(x)},
\end{eqnarray}
with the effective region $(-\infty,\infty)$, i.e.,  $x\rightarrow1$, $x_{*}\rightarrow-\infty$ and $x\rightarrow x_{h}$, $x_{*}\rightarrow \infty$, where $x_{h}$ is the location of the horizon. The effective potential $V(x)$ is 
\begin{equation}\label{eq-potential1}
    V(x)=\frac{f(x)}{\eta^2}\frac{1}{\sqrt{\Psi(x)}}\left(f(x)\sqrt{\Psi(x)}'\right)'+l(l+1)f(x),
\end{equation}
where the prime stands for the derivative with respect to $x$.  

\begin{figure*}[htbp]
\centering
\includegraphics[width=0.32\textwidth]{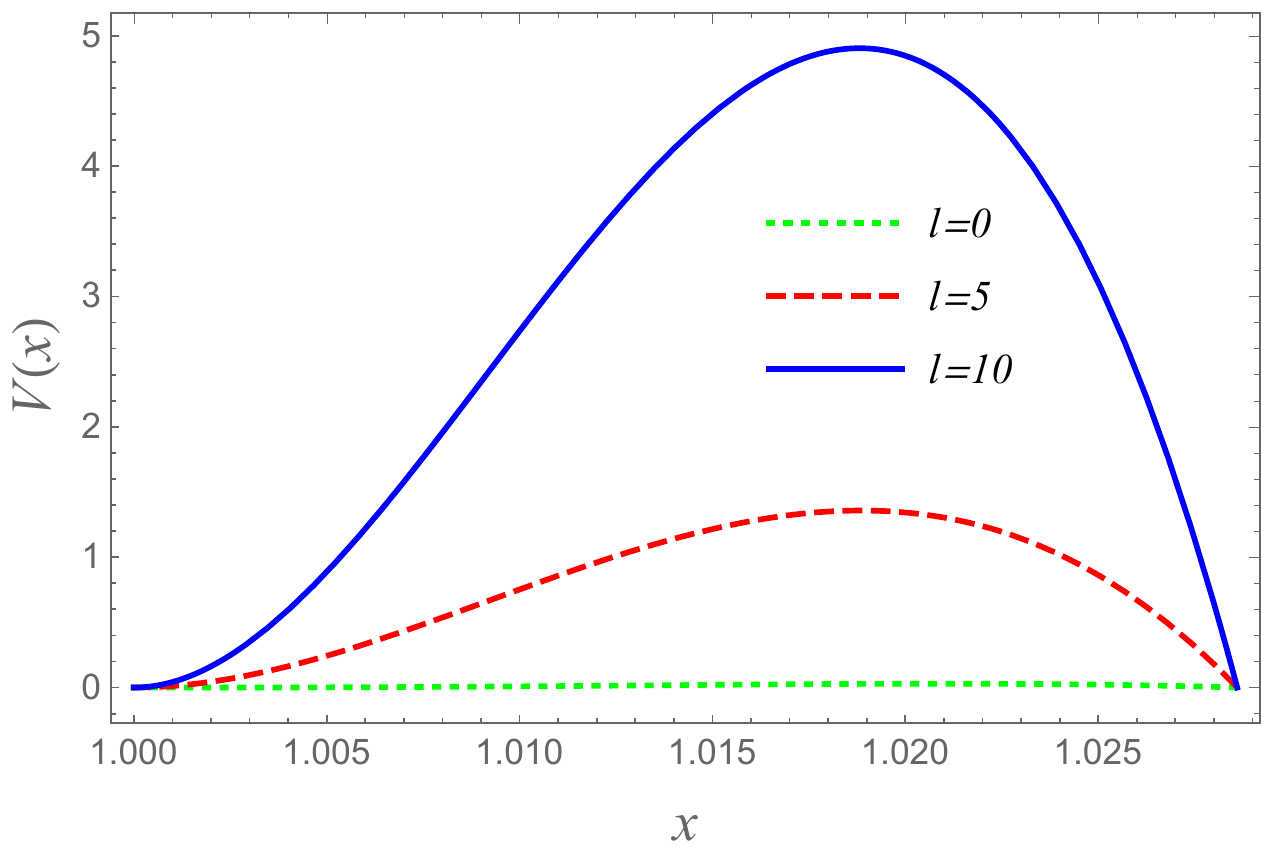}
\includegraphics[width=0.33\textwidth]{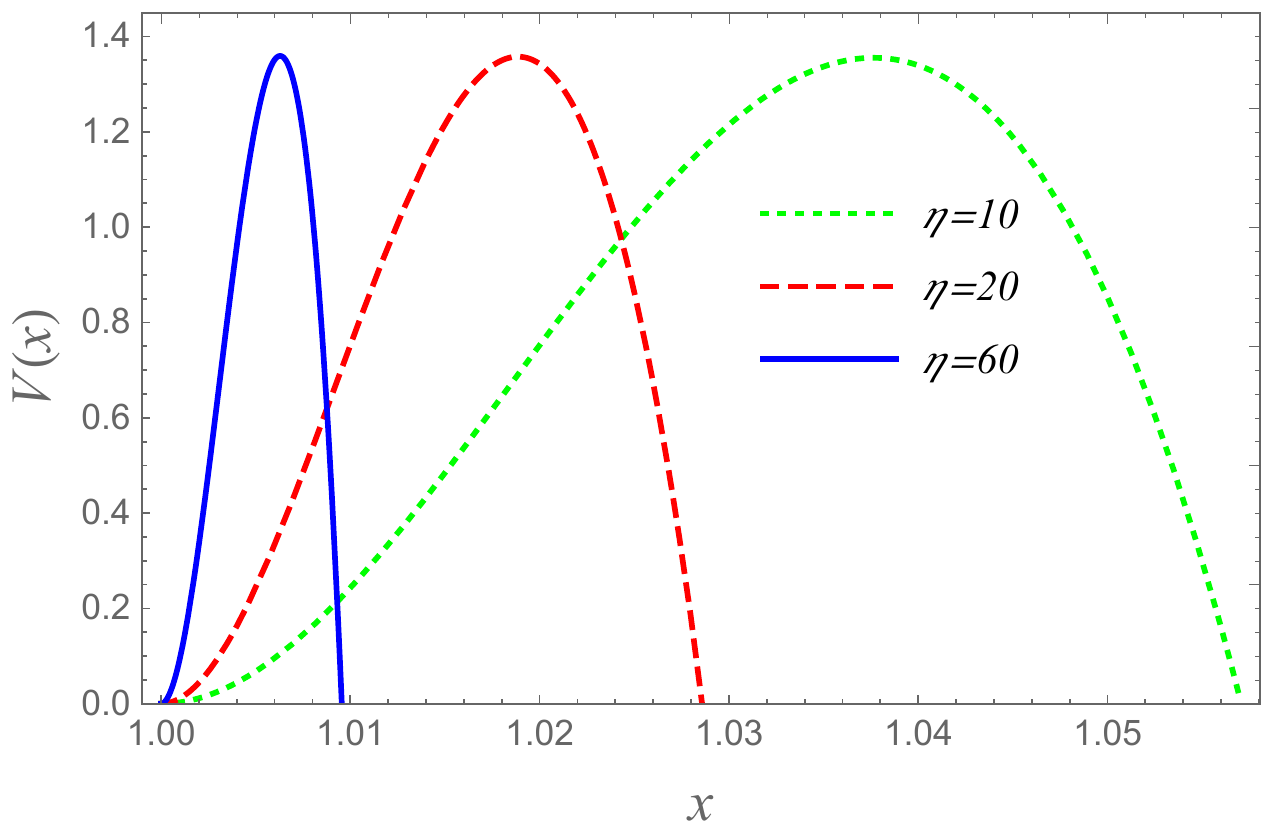}
\includegraphics[width=0.32\textwidth]{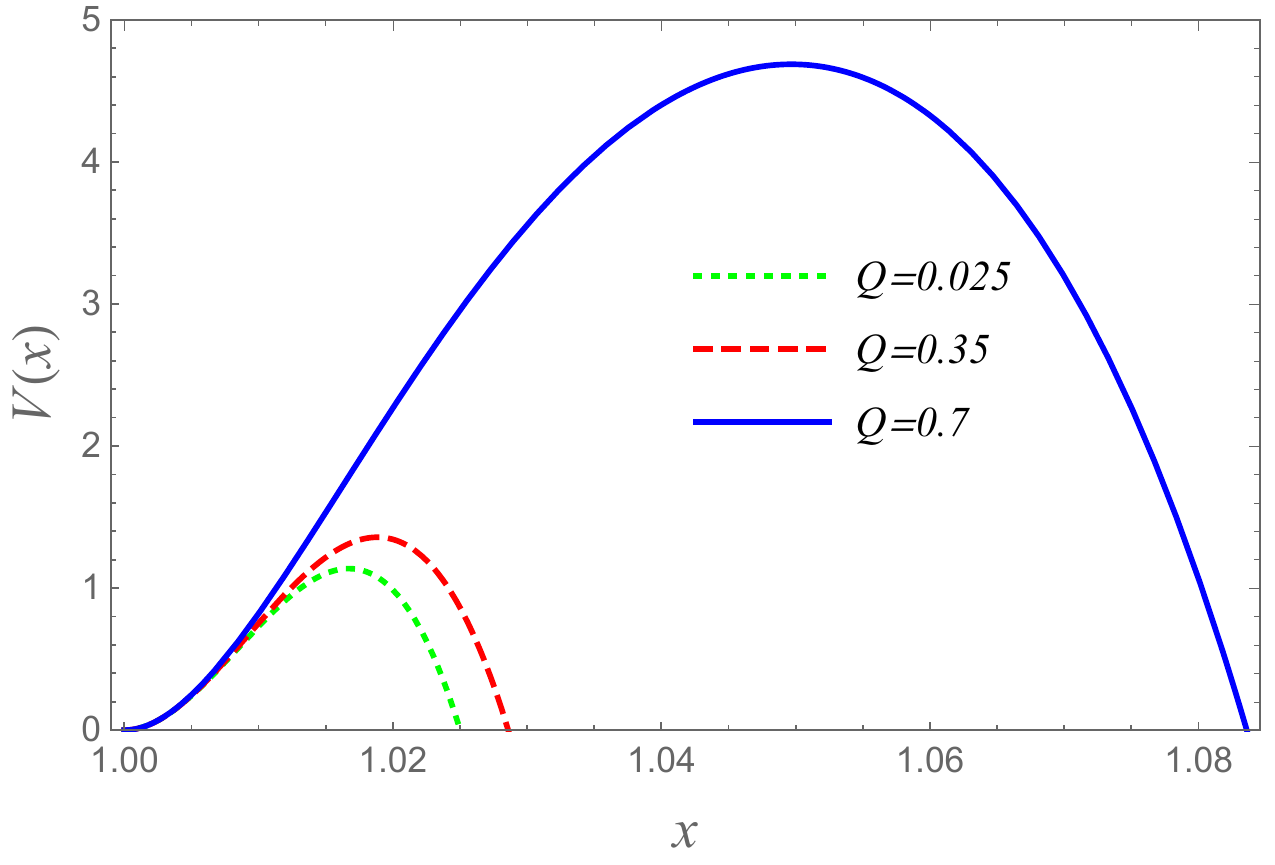}
\caption{The behaviors of the effective potential $V(x)$ with different parameters. In each {subfigure}, we fix two parameters in $(Q,\,\eta,\,l)$ and vary the third one. (Left) $Q=0.35,\, \eta=20$; (Middle) $Q=0.35,\,l=5$; (Right) $\eta=20,\,l=5$.}
\label{fig-potential}
\end{figure*}

\begin{figure}[htbp]
\centering
\includegraphics[width=0.45\textwidth]{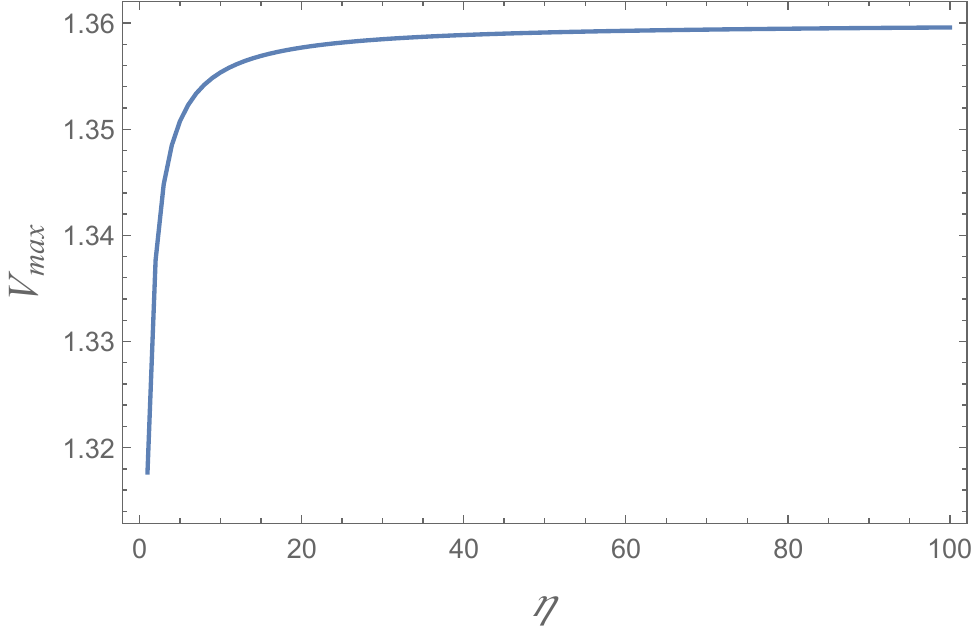}
\caption{The maximum $V_{max}$ of $V(x)$ as a function of $\eta$ for $Q=0.35,\,l=5$. }
\label{fig-potential2}
\end{figure}

Fig. \ref{fig-potential} shows the behaviors of $V(x)$ with selected parameter options. However, we verified that such behaviors are universal for parameter options with large enough $\eta$. As expected, for fixed $Q$ and $\eta$, the maximum of $V(x)$ increases as $l$ increases, while the maximum of $V(x)$ also increases as $Q$ increases if $\eta$ and $l$ are fixed. More interestingly, one may note that for the selected $Q$ and arbitrary $l$ the maximum of $V(x)$ is almost independent of the large enough parameter $\eta$. In fact, for all $Q<Q_{e}$ where the extreme value $Q_{e}$ is restricted by (\ref{eq-mass}, \ref{eq-charge}) and (\ref{eq-restricted}), the maximum of $V(x)$ increases as $\eta$ increases and reaches a constant, see Fig. \ref{fig-potential2}. This feature is consistent with Fig. \ref{fig:eta}.

\begin{figure*}[htbp]
\centering
\includegraphics[width=0.336\textwidth]{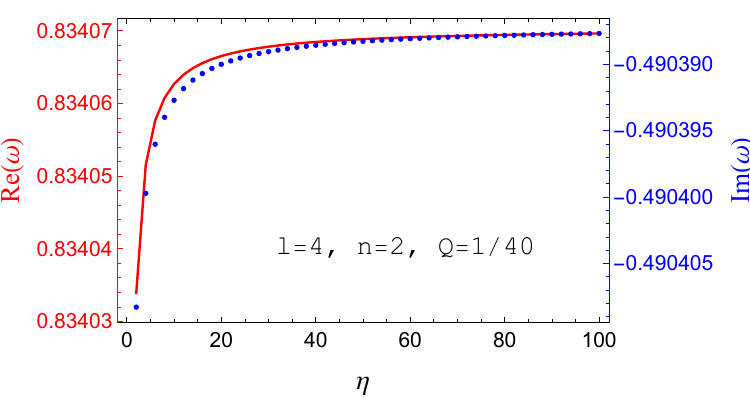}
\includegraphics[width=0.314\textwidth]{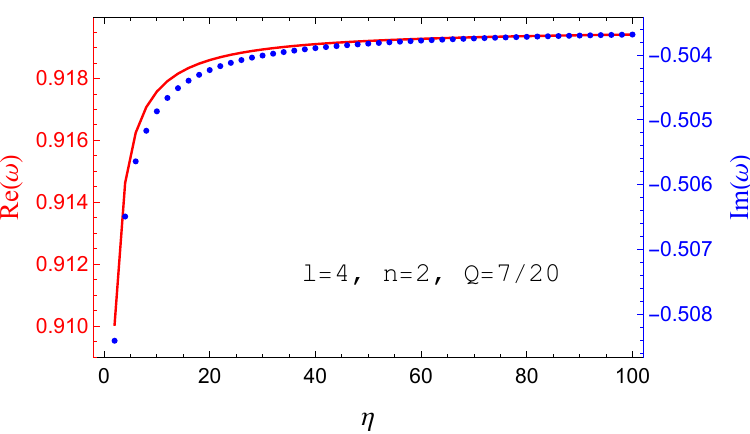}
\includegraphics[width=0.300\textwidth]{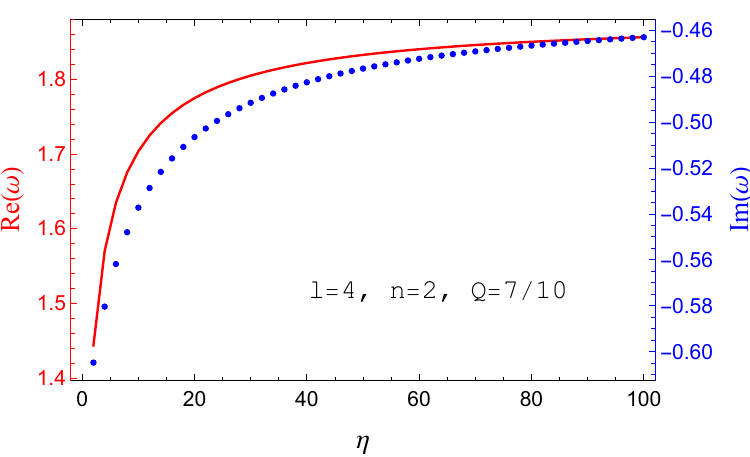}
\caption{The behaviors of $\text{Re}(\omega)$ (red solid line) and $\text{Im}(\omega)$ (blue dotted line) as functions of $\eta$ for $l=4, n=2$. (Left) $Q=0.025$; (Middle) $Q=0.35$; (Right) $Q=0.7$.}
\label{fig-nl}
\end{figure*}

To compute the QNMs one inevitably has to apply the semi-analytical or numerical methods, see the comprehensive reviews \cite{Kokkotas1999lrr,Berti2009cqg,Konoplya2011rmd}. We will adopt the so-called Hatsuda method, i.e., the WKB approximations with the Borel summation, developed in \cite{Bender1969pr,Sulejmanpasic2018CPC,Hatsuda2020prd} to calculate the QNMs, and {verify} with the higher-order WKB approximations with the Pad\'e summation \cite{Matyjasek:2017psv,Matyjasek:2019eeu}. In fact, the most important aspect of the Hatsuda method is not the Borel summation itself, but the use of the powerful \texttt{BenderWu} package developed by Sulejmanpasic and \"{U}nsal \cite{Sulejmanpasic2018CPC}, which extends WKB calculations to exceptionally high orders at the cost of restricting the computational efficiency to numerical implementations rather than analytical derivations. 

\subsection{Small $l$ and $n$}

\begin{figure*}[htbp]
\centering
\includegraphics[width=0.325\textwidth]{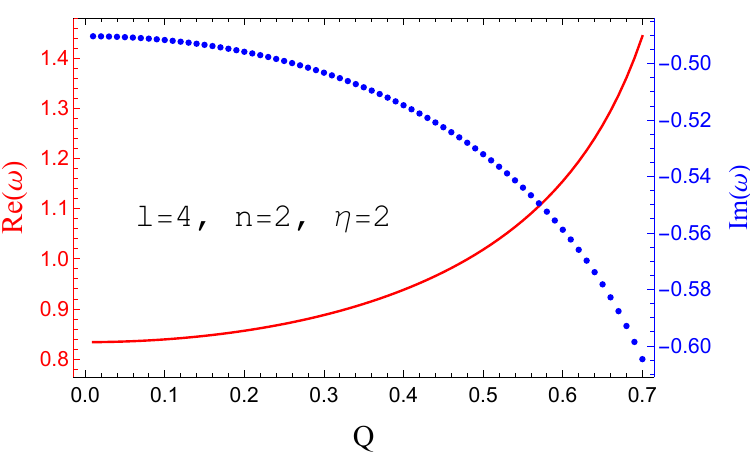}
\includegraphics[width=0.325\textwidth]{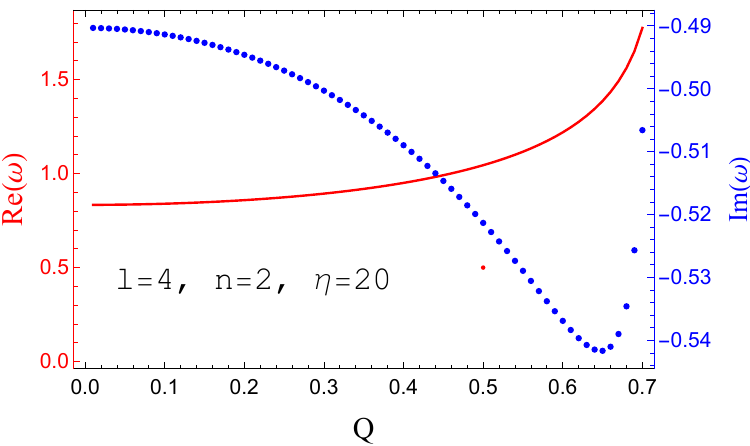}
\includegraphics[width=0.320\textwidth]{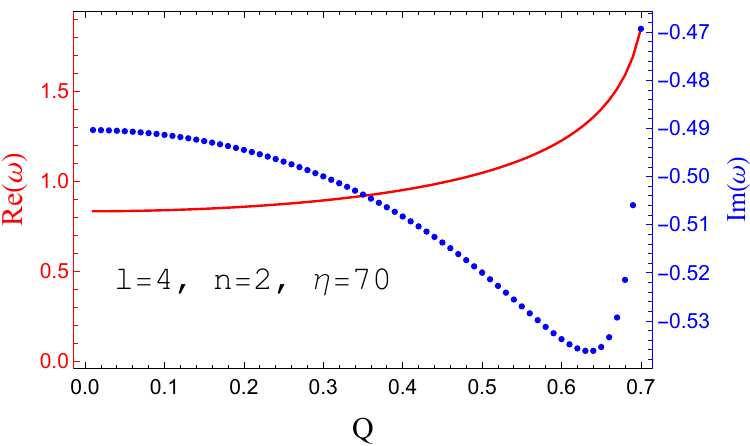}
\caption{The behaviors of $\text{Re}(\omega)$ (red solid line) and $\text{Im}(\omega)$ (blue dotted line) as functions of $Q$ for $l=4, n=2$. (Left) $\eta=2$; (Middle) $\eta=20$; (Right) $\eta=70$.}
\label{fig-nleta}
\end{figure*}

Focusing on the small $l$ and $n\leq l$, we plot the behaviors of $\text{Re}(\omega)$ and $\text{Im}(\omega)$ as functions of $\eta$ for $l=4, n=2$ and $Q=0.025, 0.35, 0.7$ in  Fig. \ref{fig-nl} (Our numerical results show that for other small $l$ and $n$ the behaviors are similar.) Such behaviors are still consistent with Fig. \ref{fig:eta}, i.e., for larger $Q$ the influences of coupling constant $\nu$ will be more obvious, and as $\eta$ approaching to infinity the results will be close to a constant. Interestingly, for arbitrary $Q<Q_e$ $\text{Re}(\omega)$ is always a monotone increasing function of $\eta$ or $Q$, while $\text{Im}(\omega)$ is no longer a monotonic function of $Q$ when $\eta$ is large enough, see Fig. \ref{fig-nleta}. In fact, for $Q\rightarrow0$ one has $\nu\rightarrow1$, which indicates that the spacetime becomes a Schwarzschild one with $\omega=0.833747-0.490338 i$ for $l=4, n=2$ scalar perturbation \footnote{To verify this, using the 50th order WKB with the Pad\'{e} summation the result becomes $\omega=0.833692-0.490325 i$, which is additionally confirmed by the continued fraction method.}. This result is agreement with the result presented in Ref.\cite{Mamani:2022akq} but differs from this in Ref. \cite{SHU2005340PLB}. Meanwhile, this result is irrelevant to $\eta$. 

Mathematically, for small enough $\eta$ Eq. \ref{eq-nu} suggests that $\nu\sim 1$, leading to $\gamma\rightarrow\infty$ and vanishing $\phi, V(\phi)$. In such case the spacetime mimics a "hairy" Reissner-Nordstr\"{o}m black hole spacetime. With the Hatsuda method and confirmed with the higher WKB method with the Pad\'{e} summation the frequency of the scalar perturbation for a RN black hole with $Q=0.7, l=4, n=2$ is $\omega=0.9257480-0.4983284 i$ while for the metric (\ref{metric}) with $Q=0.7, l=4, n=2, \eta=0.1$ one has $\omega=0.9222964-0.5349666i$. For large enough $\eta$, one has $\nu\sim4 Q^2\eta/M$ thus $\gamma\sim1$, the spacetime becomes the low energy limit of the string theory. To check this, considering $Q=0.7, l=3, n=0, \eta=400$, the frequency of the scalar perturbation is then $\omega=1.451472-0.089668i$. This result is quite close to the one obtained through the higher order WKB approximations with the Pad\'{e} summation using the exact metric given in Ref. \cite{Holzhey:1991bx}, i.e., $\omega=1.4556911-0.0858381 i$. Note if we set $M=1$ then the charge $Q=0.7$ in our case is same with the charge $Q=1.4$ in Ref. \cite{Holzhey:1991bx}.

To close this subsection we argue that the behaviors of the $\text{Im}(\omega)$ is similar to $-(n+1/2)\lambda$, where $\lambda$ is {Lyapunov exponent} given by Eq.(\ref{eq-lambda}). 

\subsection{The eikonal/geometric optics approximation: $l\gg1$}

We now consider the case $l\gg 1$, i.e., the eikonal limits. For this purpose we work on the third order WKB expansion which is good enough for large $l$ expansion. Under this approximation one has 
\begin{equation}
    \frac{i(\omega^2-V(x_{*0}))}{\sqrt{-2V''(x_{*0})}}-\Lambda_{2}(n)-\Lambda_{3}(n)=n+\frac{1}{2}, \label{eq-WKB3}
\end{equation}
where $x_{*0}$ is the solution of $V'(x_*)=0$, and the prime is the derivative with respect to $x_*$. The exact expressions of the correct terms $\Lambda_{2}(n), \Lambda_{3}(n)$ can be found in Refs. \cite{Iyer:1986np,Matyjasek:2017psv,Konoplya:2019hlu} \footnote{However, one should note that the expression in \cite{Iyer:1986np} should be slightly modified.}. By substituting the expressions of $V(x_*), \Lambda_{2}(n), \Lambda_{3}(n)$ and expanding Eq. \ref{eq-WKB3} to infinity about $l$, we can obtain
\begin{eqnarray}
    \omega&=&\omega_{R}-i\omega_{I}\nonumber \\
    &\simeq&l\sqrt{f(x_{ph})}-i(n+1/2)\sqrt{-\frac{f(x_{ph})f''(x_{ph})}{2\eta^{2}}}.\label{eq-WKBomega}
\end{eqnarray}

\begin{figure*}[htbp]
\centering
\includegraphics[width=0.4\textwidth]{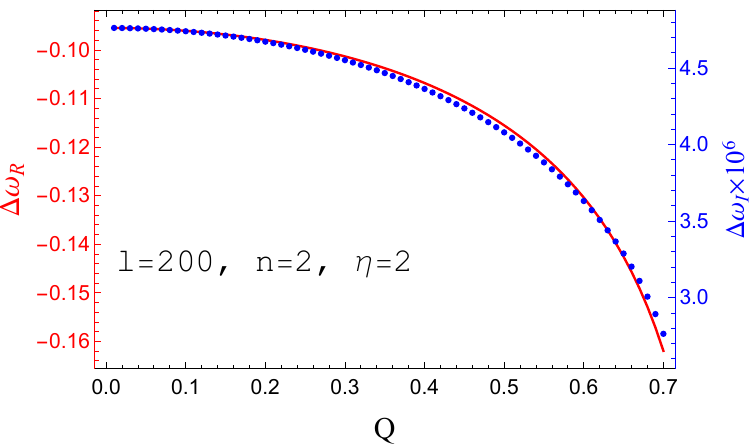}
\includegraphics[width=0.4\textwidth]{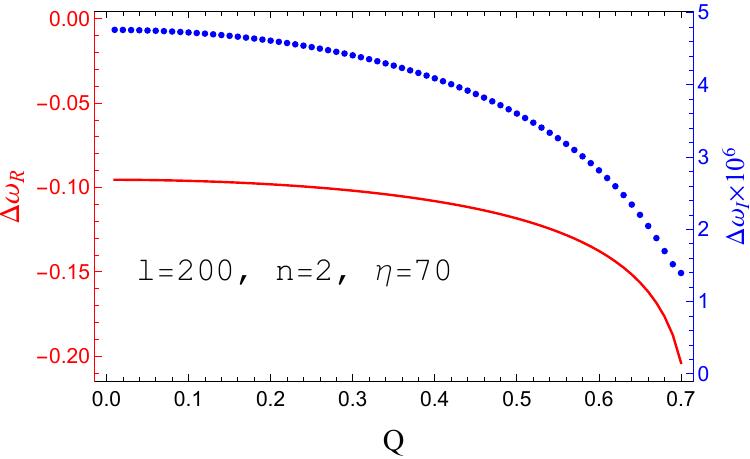}
\caption{The behaviors of $\Delta\omega_{R}$ (red solid line) and $\Delta\omega_{I}$ (blue dotted line) as functions of $Q$ for $l=200, n=2$. (Left) $\eta=2$; (Right) $\eta=70$.}
\label{fig-nldiff}
\end{figure*}

Here we have replaced $x_{*0}$ with $x_{ph}$, and the prime is now the derivative with respect to $x$. In fact, from Eqs. (\ref{eq-cng2}, \ref{eq-potential1}) we have for $l\rightarrow\infty$
\begin{eqnarray}
    \frac{dV(x_*)}{dx_*}=\frac{f(x)}{\eta}\frac{dV(x)}{dx}\approx \frac{f(x)}{\eta}l^{2}\frac{df(x)}{dx},
\end{eqnarray}
so $x_{*}=x_{*0}$ corresponds to $x=x_{ph}$. Following Eqs. (\ref{eq-lambda}, \ref{eq-omegac}) one can easily check that $\lambda\simeq\omega_{I}/(n+1/2)$ and $l\Omega_{c}\simeq\omega_{R}$. Based on the Hatsuda method we plot the deviations $\Delta\omega_{R}=l\Omega_{c}-\omega_{R}$ and $\Delta\omega_{I}=(n+1/2)\lambda-\omega_{I}$ as functions of the charge $Q$ in Fig. \ref{fig-nldiff}. Interestingly, the deviation of the real part is relatively much larger than the imaginary part, and $|\Delta\omega_{R}|$ increases with $Q$ but $\Delta\omega_{I}$ decreases with $Q$. To be more precise, the second line of Eq. (\ref{eq-WKBomega}) should be modified as 
\begin{eqnarray*}
    \omega\simeq \left(l+\frac{1}{2}\right)\sqrt{f(x_{ph})}-i(n+1/2)\sqrt{-\frac{f(x_{ph})f''(x_{ph})}{2\eta^{2}}},
\end{eqnarray*}
with such a correction the deviation $\Delta\omega_R$ between the numerical Hatsuda method and the eikonal limit will become much smaller. In addition, as $\eta$ increasing, the deviation $\Delta\omega_{R}$ stays almost same but the region of $\Delta\omega_{I}$ becomes large, see the vertical axis in Fig. \ref{fig-nldiff}. 

\section{Summaries and remarks}

In this article we have revisited the QNMs of the scalar perturbations for a general family of exact asymptotically flat hairy electrically charged black hole solutions with a non-trivial dilaton field potential. The most general metric we adopted in this article includes a conformal factor and a constant $\eta$ of integration of gravitational field introduced to obtain a dimensionless radial coordinate. Combining with another conserved charge $q$ one can find a relation between $\eta, q$ and $\nu$ thus $\eta$ can be used to govern the transition between different physical regimes. In the limit $\eta \to 0$, the spacetime action reduces to that of Einstein-Maxwell theory, and the corresponding solution becomes a diatonic Reissner-Nordström black hole. On the other hand, when $\eta \to \infty$, the parameter $\nu$ also diverges, and the action approaches the low-energy effective description of string theory.

We first computed the shadow characteristics of the asymptotically flat hairy
black holes with a dilaton potential. Explicit expressions for the Lyapunov exponent $\lambda$ and the coordinate angular velocity $\Omega_{c}$ were derived as functions of $\eta$, which provide insight into the behavior of null geodesics around the horizon.

Using the Hatsuda method and the higher order WKB approximation, we further calculated the QNMs for such kind of black holes. For small angular momentum number $l$, we observed that as the charge $Q$ increases, the real part of the QNMs increases monotonically, while the imaginary part exhibits more complex behavior—it first decreases and then increases with increasing $\eta$. Notably, as $Q$ approaches its extremal value $Q_{e}$ and the spacetime transitions to the form as the low-energy limit of string theory, the QNM spectrum coincides with the well-known solution for coupling constant equal to unity.

For large $l$, we demonstrated that both the real and imaginary parts of the QNMs are indeed consistent with the values of $\lambda$ and the coordinate angular velocity $\Omega_{c}$. Interestingly, we found that the discrepancy in the real part is significantly larger than that in the imaginary part, due to the lack of a $\eta$-independent term. Moreover, while the discrepancy in the real part increases as $\eta$ increases, the error in the imaginary part decreases with $\eta$.

These results provide a deeper understanding of the geometric and dynamical properties of black holes in this class of EMD theories, and highlight the rich structure of their perturbative stability and high-frequency behavior across different coupling regimes. However, our current analysis has been restricted to linear scalar perturbations. A natural extension would be to investigate whether analogous behavior manifests in the electromagnetic and gravitational fields perturbations.

\begin{acknowledgments}
This work is partly supported by Natural Science Foundation of China under Grants No.12375054. J. M. was partially supported by Grant No. 2022/45/B/ST2/00013 of the National Science Center, Poland. Y-Z. L. is also supported by the research funds No. 2055072312. 
\end{acknowledgments}

\nocite{*}
\bibliography{ref}

\end{document}